\documentclass[preprint,amsmath,amssymb]{revtex4}
\usepackage{dcolumn}
\usepackage{bm}
\usepackage{graphicx}
\usepackage{subfigure}
\usepackage{color}

\newcommand{\be}{\begin{eqnarray}}
\newcommand{\ee}{\end{eqnarray}}

\begin{document}

\title{Island may not save the information paradox of Liouville black holes}

\author{Ran Li$^{1,2}$}
\thanks{liran@htu.edu.cn}

\author{Xuanhua Wang$^3$}
\thanks{xuanhua.wang@stonybrook.edu}

\author{Jin Wang$^{2,3}$}
\thanks{Corresponding author: jin.wang.1@stonybrook.edu}

\affiliation{
 $^1$ Department of Physics, Henan Normal University, Xinxiang 453007, China\\
 $^2$ Department of Chemistry, SUNY at Stony Brook, Stony Brook, NY 11794, USA\\
 $^3$ Department of Physics and Astronomy, SUNY at Stony Brook, Stony Brook, New York 11790, USA}

\begin{abstract}
By using the quantum extremal island formula, we perform a simple calculation of the generalized entanglement entropy of Hawking radiation from the two dimensional Liouville black hole. No reasonable island was found when extremizing the generalized entropy. We explain qualitatively the reason why the page curve cannot be reproduced in the present model. This suggests that the islands may not necessarily save the information paradox for the Liouville black holes. 
\end{abstract}

\maketitle

\section{Introduction}\label{sect:intro}

Recently, a remarkable progress was made in studying the information paradox of the black holes, which is caused by the Hawking radiation \cite{Hawking:1976ra}. It was shown that the island proposed to be in the entanglement wedge of the radiation should be taken into account when calculating of entanglement entropy of Hawking radiation \cite{Penington:2019npb,Almheiri:2019psf,Almheiri:2019hni,Almheiri:2019yqk}. It was proposed that the entanglement entropy of the Hawking radiation should be given by \cite{Almheiri:2019yqk}
\begin{align}
\label{formula}
S_R={\rm min}\left\{ {\rm ext} \left[\frac{A[\partial I]}{4G_N}+S_{\rm bulk}[{\rm Rad}\cup I]\right]\right\} \,,
\end{align}
where "Rad" is the region that the distant observer collects the radiation and $I$ is the island.
$S_{\rm bulk}[{\rm Rad}\cup I]$ is the entanglement entropy of the quantum fields (including gravitational field) in the region ``Rad'' and island $I$. $\partial I$ is the boundary surface of the island and ${A[\partial I]}/{4G_N}$ is the area of the surface. The terms in the square brackets are just the generalized entropy of Bekenstein \cite{Bekenstein:1974ax}. If extremizing the generalized entropy, one can determine the location of the island. The minimum value is chosen as the entanglement entropy of the Hawking radiation if there are more than one extremum. It was shown that the Page curve \cite{Page:1993wv} can be reproduced by using this formula. See \cite{Almheiri:2020cfm} for a nice review of this formula.

The island formula was initially proposed for the AdS black holes from the holographic perspective. However, subsequent works show that this formula does not require the AdS/CFT correspondence. The island formula was justified by the replica wormholes in the gravitational path integral \cite{Penington:2019kki,Almheiri:2019qdq,Goto:2020wnk}.
In particular, it may also be applicable to asymptotically flat black holes. Although the island formula has been successfully applied to many types of black hole spacetime, including black holes in JT gravity, higher dimensional asymptotically flat and AdS black holes, and the Page curves are properly reproduced [12-43], it is still interesting to check whether this formula is applicable to other types of black hole solutions.

In this note, we investigate whether the quantum extremal island formula can be applied to resolve the information paradox of the eternal Liouville black hole in the two dimensional dilaton gravity with the exponential potential. This type of model was initially introduced by R. B. Mann \cite{Mann:1993rf}. Later, this model was also investigated by J. Cruz et al in \cite{Cruz:1997nj,Cruz:1996pg}. It was shown that this model has the analytical black hole solutions. The black hole solution in this model was shown to have Hawking temperature proportional to its mass. This implies that the black hole will never evaporate completely consistent with the third law of the thermodynamics. We will show that the entropy of the Hawking radiation without islands grows with time linearly, which contradicts the finite degrees of freedom of black hole. Furthermore, we demonstrate that there seems no reasonable quantum extremal surface when extremizing the generalized entanglement entropy by using the island formula. We explain qualitatively the reason why the page curve cannot be reproduced in the present model. This implies that the island formula may not necessarily save the information paradox for the Liouville black holes.

\section{Liouville black holes in 2D dilaton gravity}\label{Sect:LBH}

We consider the model of 2D dilaton gravity with the exponential potential, the action of which is explicitly given by \cite{Mann:1993rf,Cruz:1997nj,Cruz:1996pg}
\be\label{action}
I=\frac{1}{2\pi} \int d^2x \sqrt{-g}\left(\phi R+4\lambda^2 e^{\beta\phi} \right)\;.
\ee
Note that the signs of the parameters $\lambda^2$ and $\beta$ will be determined in the following. This model is similar to the deformed JT gravity recently investigated by Witten et.al \cite{Witten:2020ert,Witten:2020wvy,Momeni:2020zkx,Alishahiha:2020jko,Kim:2020eqn}. However, we are considering the asymptotically flat black hole solutions while the asymptotically AdS black holes have been studied in \cite{Witten:2020ert,Momeni:2020zkx,Kim:2020eqn}.

In conformal gauge, the spacetime metric can be written in terms of the so-called Kruskal coordinates as
\be
ds^2=-e^{2\rho(x^+,x^-)}dx^+ dx^-\;.
\ee
The equation of motion of the model can be given by
\be
\partial_+\partial_-\left(\rho-\frac{\beta}{2}\phi\right)&=&0\;,\\
\partial_+\partial_-\left(\rho+\frac{\beta}{2}\phi\right)&=&-\lambda^2\beta e^{2\left(\rho+\frac{\beta}{2}\phi\right)}\;,\label{liouville}\\
\label{constraint}
-\partial_{\pm}^2\phi+2\partial_{\pm}\rho\partial_{\pm}\phi&=&0\;.
\ee
It can be seen that the field $\rho-\frac{\beta}{2}\phi$ is a free field and $\rho+\frac{\beta}{2}\phi$ satisfies the Liouville equation. It was shown that by setting $\rho=\frac{\beta}{2}\phi$, Eq.(\ref{liouville}) can be rewritten as
\be
\partial_+\partial_-\left(2\rho\right)=-\lambda^2 \beta e^{4\rho}\;.
\ee
This is exactly the liouville equation, and the general solution is given by
\be\label{generalsol}
\rho=\frac{1}{4} \ln\frac{\partial_+F\partial_-G}{
\left( 1+\lambda^2\beta FG \right)^2} \;,
\ee
where $F(G)$ is an arbitrary function of $x^+(x^-)$. Using the constraint equation Eq.(\ref{constraint}), one can get
\be
\partial_{\pm}^2\rho=2\left(\partial_{\pm}\rho\right)^2\;.
\ee
By substituting Eq.(\ref{generalsol}) into the above equation, one can reach the general solution as \cite{Mann:1993rf,Cruz:1997nj,Cruz:1996pg} 
\be
e^{2\rho}=\frac{1}{Cx^+x^-+Ax^++Bx^-+D}\;,
\ee
where $A$, $B$, $c$, and $D$ are arbitrary constant satisfying the constraint 
\be
AB-CD=-\lambda^2 \beta\;. 
\ee
For simplicity, we only consider the solution with $A=B=0$, which can also be treated as the gauge fixing of Kruskal coordinates $x^+\rightarrow x^+-\frac{A}{C}$ and $x^-\rightarrow x^--\frac{B}{C}$.  In this case, the solution can be written as 
\be\label{bhsolution}
e^{-2\rho}=e^{-\beta\phi}=\frac{\lambda^2\beta}{C}+Cx^+ x^-\;.
\ee

\begin{figure}
  \centering
  \includegraphics[width=10cm]{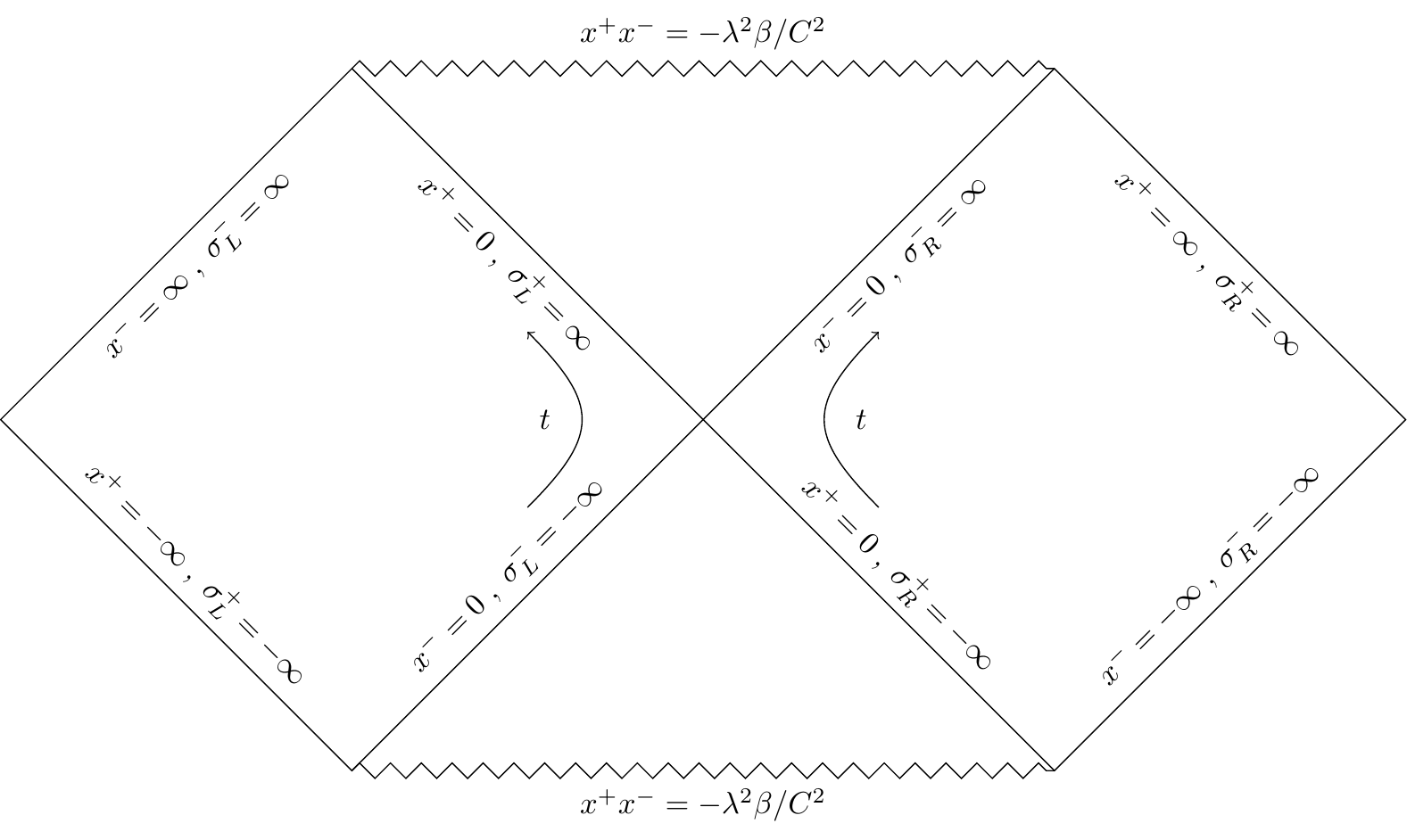}\\
  \caption{Penrose diagram of the Liouville black holes.}
  \label{PenroseDiagram}
\end{figure}

This solution represents a black hole solution when $C<0$ and $\lambda^2\beta<0$. The horizon is located at
\be
x^+x^-=0\;.
\ee
The scalar curvature is given by
\be
R=-4\lambda^2\beta e^{\beta\phi}=\frac{-4\lambda^2\beta}{\frac{\lambda^2\beta}{C}+Cx^+ x^-}\;.
\ee 
Therefore, the curvature singularity is located at
\be
\frac{\lambda^2\beta}{C}+Cx^+ x^-=0\;.
\ee
One can find that the scalar curvature blows up at the singularity and approaches to zero asymptotically. The solution is also asymptotically flat. The Penrose diagram is shown in Fig.(\ref{PenroseDiagram}), which is the same as that of the Schwarzschild black hole.

We now explain how to determine the signs of the parameters. If $C>0$, the solution (\ref{bhsolution}) does not have the appropriate metric signature in the $x^+x^-<0$ region outside the horizon. If $\lambda^2\beta>0$, the location of the curvature singularity $x^+x^-=-\frac{\lambda^2\beta}{C^2}<0$, i.e. the singularity is outside the horizon, which means that the singularity is naked. Therefore, assuming the cosmic censorship conjecture, we should consider the black hole solution with the constraint conditions $C<0$ and $\lambda^2\beta<0$.

The asymptotic flatness of the black hole is also manifest in the coordinates introduced in the following. In the right region of Penrose diagram, by introducing the coordinate transformation
\be
\sqrt{|C|}x^{\pm}=\pm e^{\pm\sqrt{|C|}\sigma^{\pm}_{R}}\;,
\ee
the metric can be written as
\be
ds^2=\frac{-d\sigma^+_R\sigma^-_R}{1+\frac{\lambda^2\beta}{C}
e^{-\sqrt{|C|}\left(\sigma^+_R-\sigma^-_R\right)}}\;.\label{rightmetric}
\ee
Obviously, when $\sigma^+_R-\sigma^-_R\rightarrow\infty$, the spacetime approaches to the flat one.
In the left region, by introducing the coordinate transformation
\be
\sqrt{|C|}x^{\pm}=\mp e^{\mp\sqrt{|C|}\sigma^{\pm}_{L}}\;,
\ee
the metric can be written as
\be
ds^2=\frac{-d\sigma^+_L\sigma^-_L}{1+\frac{\lambda^2\beta}{C}
e^{\sqrt{|C|}\left(\sigma^+_L-\sigma^-_L\right)}}\;.
\ee
When $\sigma^+_L-\sigma^-_L\rightarrow-\infty$, the spacetime also approaches to the flat one. Therefore, similar to the eternal Schwarzschild black holes, there are two asymptotically flat universes in the Penrose diagram. Notice that $\sigma^{\pm}=t\pm\sigma$ holds both in the left region and in the right region. By definition, the time $t$ in the two regions are all upward. This is essential for the information paradox of the eternal black holes.

Introducing the coordinate transformation 
\be
1+\frac{\lambda^2\beta}{C}e^{-2\sqrt{|C|}\sigma}=\frac{1}{1-e^{-2\sqrt{|C| x}}}\;,
\ee
the metric (\ref{rightmetric}) can be written as 
\be\label{schmetric}
ds^2=-\left(1-e^{-2\sqrt{|C| x}}\right)dt^2+\frac{1}{1-e^{-2\sqrt{|C| x}}}dx^2\;.
\ee
It can be shown that the time $t$ has the periodicity along the imaginary axis. Introducing a new spatial coordinate defined by 
\be
x=\frac{\sqrt{|C|}}{2} R^2\;,
\ee
near the horizon $x=0$, the metric (\ref{schmetric}) in the Euclidean time $t=i\tau$ coordinate has the form of 
\be
ds^2=R^2d\left(\sqrt{|C|}\tau\right)^2+dR^2\;.
\ee
Clearly, the Euclidean time has the periodicity of $\frac{2\pi}{\sqrt{|C|}}$. Hawking temperature, which is the inverse of the period, is given by
\be
T=\frac{\sqrt{|C|}}{2\pi}\;.
\ee

The mass of the black hole is the conserved charge associated with the Killing vector $k^\mu=\left(\frac{\partial}{\partial t}\right)^\mu=(1,0)$. The Noether charge is given by \cite{Cruz:1997nj,Cruz:1996pg}
\be
Q=\frac{1}{2\pi}\epsilon^{\mu\nu} 
\left.\left(2k_\mu \nabla_\nu \phi+\phi \nabla_\mu k_\nu\right)\right|_{x=+\infty}\;.
\ee
By employing the expression of the dilaton field in the $(t, x)$ coordinates
\be
\phi=-\frac{2}{\beta} \sqrt{|C|} x -\frac{1}{\beta} \ln \frac{\lambda^2\beta}{C}\;,
\ee
one can easily calculate the mass of the black hole, which is given by 
\be
M=\frac{2\sqrt{|C|}}{\beta\pi}\;.
\ee

The expression of the black hole mass gives another constraint for the parameters. It is clear that $\beta$ must be positive in order to have a reasonable conserved quantity for the black hole. Therefore, we conclude that our full restrictions for the parameters are 
\be\label{restrictions}
C<0\;,\;\;\beta>0\;,\;\;\lambda^2<0\;.
\ee
In the following, we will work with these conditions to calculate the entropy of the Hawking radiation.

It can be seen that the Hawking temperature is proportional to the mass as
\be
T=\frac{\beta}{4}M\;.
\ee
This indicates that the evaporating time of this type of black hole is infinity. According to the first law of thermodynamics, the entropy of the black hole can be calculated as
\be
S_{BH}=\int\frac{dM}{T}=\frac{4}{\beta}\ln M-S_0\;,
\ee
where $S_0$ is an integral constant. The dilaton field at the horizon is
\be
\phi_H=\phi(x^+x^-=0)=\frac{1}{\beta}\ln\frac{C}{\lambda^2\beta}\;.
\ee
If choosing $S_0=\frac{2}{\beta}\ln\left(\frac{-4\lambda^2}{\beta\pi^2}\right)$, the black hole entropy is related to the dilaton field at the horizon
\be\label{BH_entropy}
S_{BH}=2\phi_H\;.
\ee
This is consistent with the conclusion that the dilaton field plays the role of area in the case of the two dimensional dilaton gravity.

The above expression of the entropy is also the Wald entropy for the two dimensional dilaton gravity. Wald entropy is defined as the Noether charge associated with diffeomorphism invariance of a theory \cite{Wald:1993nt}. Iyer and Wald derived the expression of the entropy for the two dimensional dilaton gravity, which is given by \cite{Iyer:1994ys}
\be
S_{Wald}=\frac{4\pi}{\sqrt{-g}} \frac{\partial \mathcal{L}}{\partial R} \;,
\ee
where $\mathcal{L}$ is the Lagrangian of the theory and the right hand side is evaluated on the horizon. For the action (\ref{action}), the Wald entropy is then given by Eq.(\ref{BH_entropy}).

\section{Entropy of radiation without islands}

In this section, we consider the entropy of Hawking radiation without the contribution of islands. This sharpens the information paradox for the Liouville black holes.

\begin{figure}
  \centering
  \includegraphics[width=10cm]{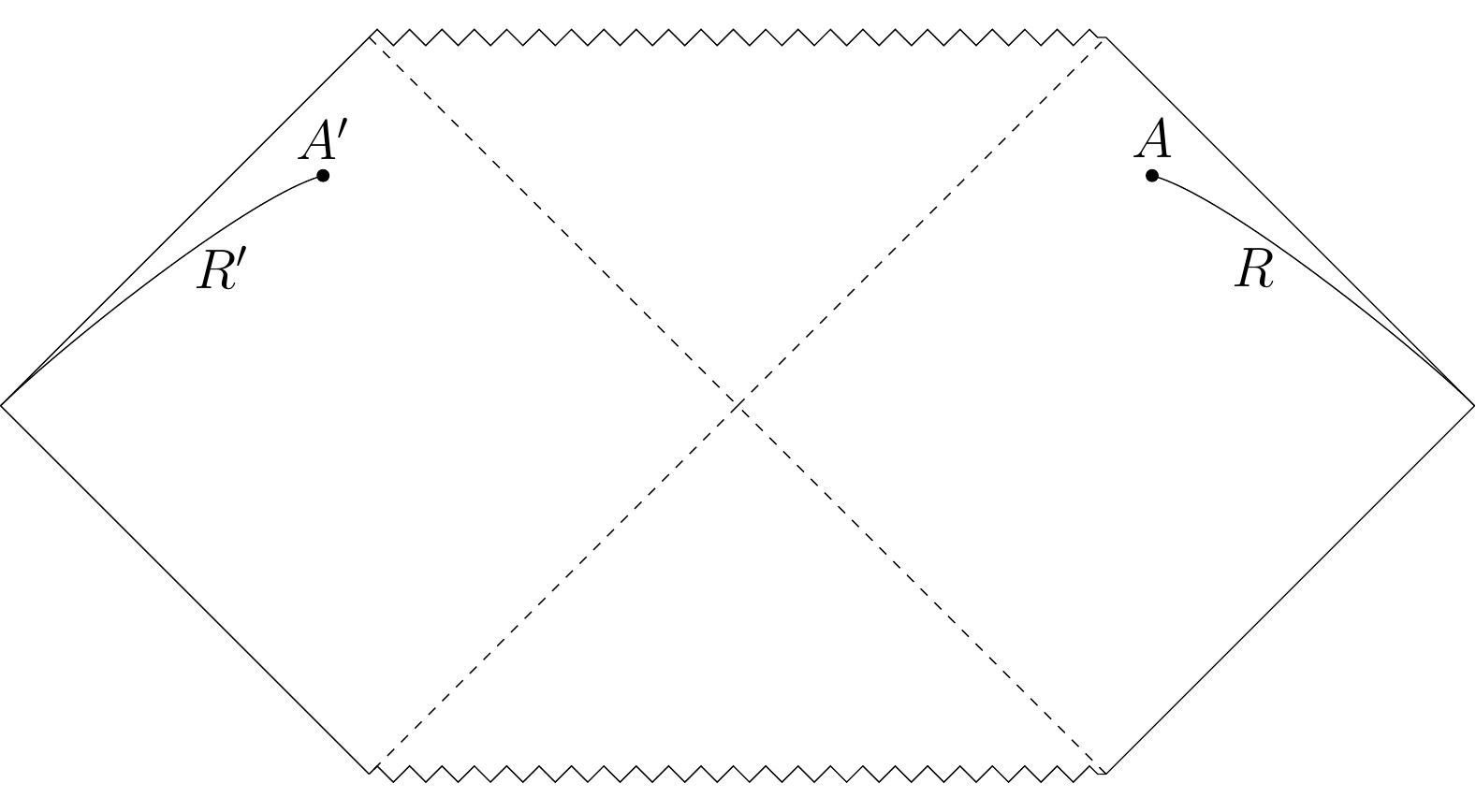}\\
  \caption{Penrose diagram without island. The Hawking radiation from the two-sided black holes is collected by the observer in the region $R'\cup R$}
  \label{PenroseDiagram_withoutisland}
\end{figure}

As shown in Fig.(\ref{PenroseDiagram_withoutisland}), the radiation is taken to be in $R$ and $R'$. The boundary points in the right and left radiation regions are denoted as $A$ and $A'$. In general, $A$ and $A'$ are assumed to be far away from the right and left horizons. In the $\sigma$ coordinates, we have
\be
\sigma_A^{\pm}=t\pm b\;,\;\;\;
\sigma_{A'}^{\pm}=t\mp b\;.
\ee
Without islands, we assume that the total system is in a pure state at $t=0$ moment. In this case, we just need to calculate the bulk entropy of the intervals $[A',A]$. We assume the conformal field is taken to be in the vacuum state in $(x^+,x^-)$ coordinates. The entanglement entropy is given by
\be
S&=&\frac{N}{6}\log \left|\left(x_A^+-x_{A'}^+\right)\left(x_A^- -x_{A'}^-\right)
e^{\rho(A)}e^{\rho(A')}\right|\nonumber\\
&=&\frac{N}{6}\log\frac{4e^{2\sqrt{|C|b}}\cosh^2\left(\sqrt{|C|}t\right)}
{|C|\left(\frac{\lambda^2\beta}{C}+e^{2\sqrt{|C|}b}\right)}\nonumber\\
&\simeq&\frac{N}{3}\sqrt{|C|}t+\cdots
\ee
where $N$ is the central charge of the bulk CFT and the last step uses the late time approximation.

This indicates that the entropy of Hawking radiation grows with time linearly as observed by the distant observer. This entropy counts the entanglement between the emitted particles and the degree of freedom behind the event horizons. However, the black hole only has finite number degrees of freedom which is counted by the Bekenstein-Hawking entropy. Therefore, we have encountered the information paradox for the Liouville black holes. We expect that the appearance of the islands at the late time may save the contradiction.

\section{Island formula in Liouville black holes}

In this section, we consider the generalized entanglement entropy of the Hawking radiation by using the quantum extremal island formula.

\begin{figure}
  \centering
  \includegraphics[width=10cm]{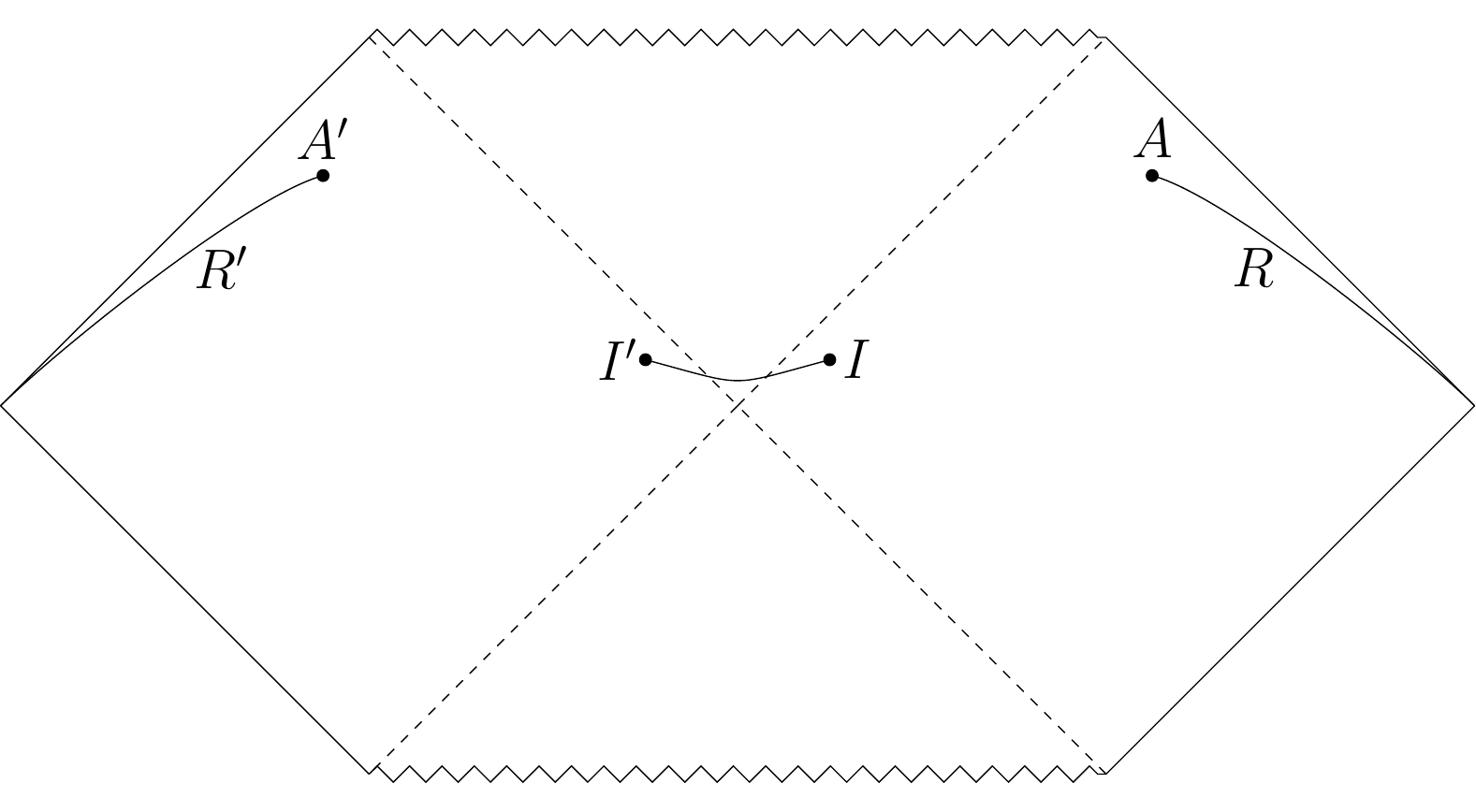}\\
  \caption{Penrose diagram with the assumed islands in the region $[I',I]$.}
  \label{PenroseDiagram_island}
\end{figure}

The Penrose diagram with the assumed islands is presented in Fig.(\ref{PenroseDiagram_island}).
We assume that the island is in the region $[I',I]$. Considering the symmetry of the left and right regions, we just need to calculate the entanglement entropy of the matter fields in the interval $[I,A]$ and the area term contribution from the island. As illustrated above, the area term is represented by the dilaton field. Therefore, the generalized entanglement entropy of the radiation taking the islands' contribution into account is given by
\be\label{generalizedentropy}
S&=&4\phi\left(x_I^+,x_I^-\right)+\frac{N}{3}\log \left|\left(x_A^+-x_{I}^+\right)\left(x_A^- -x_{I}^-\right)
e^{\rho(A)}e^{\rho(I)}\right|\nonumber\\
&=&-\frac{4}{\beta}\log\left(\frac{\lambda^2\beta}{C}+Cx_I^+ x_I^-\right)
+\frac{N}{3}\log \frac{|\left(x_A^+-x_{I}^+\right)\left(x_A^- -x_{I}^-\right)|}
{\sqrt{\frac{\lambda^2\beta}{C}+Cx_I^+ x_I^-}\sqrt{\frac{\lambda^2\beta}{C}+Cx_A^+ x_A^-}}\;.
\ee
Extremizing the generalized entropy $S$ with respect to $x_I^+$ and $x_I^-$, we have
\be\label{extremeq1}
-\frac{4}{\beta}\frac{C x_I^-}{\frac{\lambda^2\beta}{C}+Cx_I^+ x_I^-}
-\frac{N}{3}\frac{1}{x_A^+-x_I^+}-\frac{N}{6}\frac{C x_I^-}{\frac{\lambda^2\beta}{C}+Cx_I^+ x_I^-}=0\;,\\
-\frac{4}{\beta}\frac{C x_I^+}{\frac{\lambda^2\beta}{C}+Cx_I^+ x_I^-}
+\frac{N}{3}\frac{1}{x_I^--x_A^-}-\frac{N}{6}\frac{C x_I^+}{\frac{\lambda^2\beta}{C}+Cx_I^+ x_I^-}=0\;.
\label{extremeq2}
\ee

We work in the semiclassical regime, where the parameter $\beta$ and the central charge $N$ of the CFT satisfies the condition 
\be\label{semicond}
\frac{1}{\beta}\gg N \gg 1\;.
\ee
Note that $\frac{1}{\beta}$ is proportional to the mass of the black hole. Therefore, the semiclassical condition (\ref{semicond}) is valid as long as the mass is big enough. In this case, the first terms in Eq.(\ref{extremeq1}) and (\ref{extremeq2}) (which come from the classical gravity entropy) dominate over the third terms (which come from the bulk CFT  entropy). However, we still track the effects of the third terms in Eq.(\ref{extremeq1}) and (\ref{extremeq2}). It is generally believed that under the semiclassical condition (\ref{semicond}), the quantum extremal surface prescription and island formula are supposed to reproduce the Page curve of the Hawking radiation.

Combining the two equations (\ref{extremeq1}) and (\ref{extremeq2}), we can get
\be\label{relation}
x_I^+ x_A^-=x_I^- x_A^+\;.
\ee
This equation implies that the boundary of the island (if it exists) should be located at the same timeslice with the observer far from the black hole. Because $x_A^+>0$ and $x_A^-<0$, Eq.(\ref{relation}) requires that the right boundary point $I$ of the island should be located in the right region of the Penrose diagram. By substituting Eq.(\ref{relation}) back to the equations (\ref{extremeq1}) and (\ref{extremeq2}), we get
\be\label{xIeq}
\left(\frac{N}{12}-\frac{2}{\beta}\right)C \frac{x_A^+}{x_A^-}\left(x_I^-\right)^2
+\left(\frac{N}{12}+\frac{2}{\beta}\right)C x_A^+ x_I^-+\frac{N\lambda^2\beta}{6C}=0\;.
\ee

Remember that the restrictions for the parameters are 
\be
C<0\;,\;\;\beta>0\;,\;\;\lambda^2<0\;.
\ee
Obviously, $\left(\frac{N}{12}+\frac{2}{\beta}\right)C x_A^+$, which is the coefficient of the $(x_I^-)^1$ term, is negative, and $\frac{N\lambda^2\beta}{6C}$ is positive. 

Under the semiclassical condition Eq.(\ref{semicond}), $\left(\frac{N}{12}-\frac{2}{\beta}\right)$, which is the coefficient of the first term of Eq.(\ref{xIeq}), cannot be zero. However, to track the general discussion, we still distinguish the cases of $\left(\frac{N}{12}-\frac{2}{\beta}\right)\neq 0$ and $\left(\frac{N}{12}-\frac{2}{\beta}\right)= 0$. When $\left(\frac{N}{12}-\frac{2}{\beta}\right)\neq 0$, the solutions are given by
\be\label{islandsol}
x_I^-=-\frac{(N\beta+24)}{2(N\beta-24)}\left[
1\pm\left(1-\frac{8N(N\beta-24)}{(N\beta+24)^2}
\frac{\lambda^2\beta^2}{C^2}\frac{1}{x_A^+x_A^-}\right)^{1/2}\right]x_A^-\;.
\ee
In the case of large $x_A^+$, one can get two groups of solutions by expanding the solutions in Eq. (\ref{islandsol}) as the series of $\frac{1}{x_A^+}$. The first group of solution corresponding to $'-'$ in Eq.(\ref{islandsol}) is given by
\be\label{novelsolution}
(x_I^-)_1=-\frac{2N}{(N\beta+24)}\frac{\lambda^2\beta^2}{C^2}\frac{1}{x_A^+}\;,\;\;\;
(x_I^+)_1=-\frac{2N}{(N\beta+24)}\frac{\lambda^2\beta^2}{C^2}\frac{1}{x_A^-}\;,
\ee
while the second group of solution corresponding to $'+'$ in Eq.(\ref{islandsol}) is given by
\be\label{wrongisland}
(x_I^-)_2=-\left(\frac{N\beta+24}{N\beta-24}\right)x_A^-\;,\;\;\;
(x_I^+)_2=-\left(\frac{N\beta+24}{N\beta-24}\right)x_A^+\;.
\ee
Note that the higher order terms of $\frac{1}{x_A^+}$ are omitted in the above expressions. 

When $\left(\frac{N}{12}-\frac{2}{\beta}\right)=0$, it can be easily checked that the solution can also given by the first group of solutions in Eq.(\ref{novelsolution}). Therefore, we have obtained all of the solutions of the possible locations of the island, which are given by Eq.(\ref{novelsolution}) and (\ref{wrongisland}) respectively.

Now let us discuss the implications of these solutions given by Eq.(\ref{novelsolution}) and (\ref{wrongisland}). Firstly, the second group of solution given by Eq.(\ref{wrongisland}) seems to be unphysical. Under the semiclassical condition given by Eq.(\ref{semicond}), $0<N\beta\ll 1$. Therefore, we have 
\be
(x_I^-)_2> x_A^-\;,\;\;(x_I^+)_2> x_A^+\;.
\ee
In this case, the boundary point $I$ of the island is located in the region of radiation. This solution for the boundary point $I$ is not physical. 

\begin{figure}
  \centering
  \includegraphics[width=10cm]{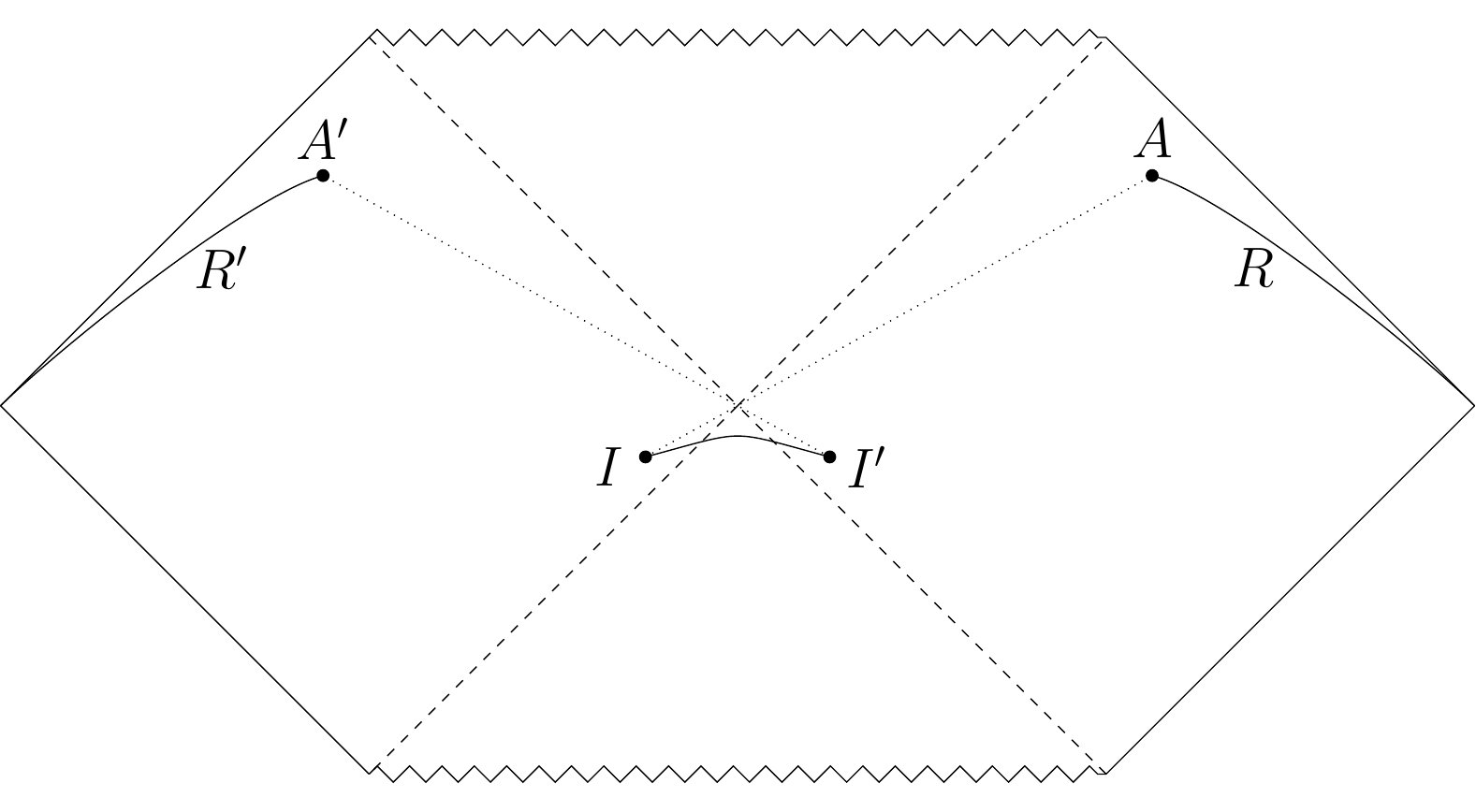}\\
  \caption{Penrose diagram with the islands which are given by the first group of solutions..}
  \label{PenroseDiagram_unphysical}
\end{figure}

Then, the first group of solution is also shown to be unreasonable. Under the restriction condition given by Eq.(\ref{restrictions}) and the semiclassical condition given by Eq.(\ref{semicond}), the prefactor of the solution (\ref{novelsolution}) satisfies 
\be
-\frac{2N}{(N\beta+24)}\frac{\lambda^2\beta^2}{C^2}>0\;.
\ee
Therefore, we have
\be
(x_I^-)_1>0\;,\;\;\;
(x_I^+)_1<0\;.
\ee
In this case, the boundary point $I$ is located in the left region near the horizon. The corresponding Penrose diagram is shown in Fig.(\ref{PenroseDiagram_unphysical}), which looks strange. It can be seen that the left boundary $I'$ and the right boundary $I$ of the island are exchanged. This will lead to a very strange entanglement wedges for the islands. This kind of solutions seems to be unphysical too. 

In summary, Eq.(\ref{relation}) requires that the right boundary surface $I$ of the island should be located in the right region of the Penrose diagram. Our simple calculation shows that, for the two dimensional Liouville black holes, there is no reasonable quantum extremal surface when extremizing the generalized entanglement entropy. This implies that the island formula may not resolve the information paradox for the Liouville black holes.

At last, we give the qualitative argument why a reasonable extremal point cannot be obtained by using the island formula. In the papers which studied the information paradox of the eternal black holes (for example, Ref.\cite{Almheiri:2019yqk} and Ref.[12-43]), the dilaton field or the area term that counts the classical gravitational contribution to the generalized entropy increases when the boundary point $I$ moves outwards from the black hole horizon, while the bulk CFT entropy between the boundary point $I$ and the observer $A$ decreases. In this case, it is possible to find the extremal point of the generalized entropy. In the present case, the dilaton field has the opposite behavior compared to the previous examples. Note that 
\be
\phi(x_I^+, x_I^-)=-\frac{1}{\beta}\log\left(\frac{\lambda^2\beta}{C}+Cx_I^+ x_I^-\right)\;.
\ee
When the boundary point $I$ moves outwards from the black hole horizon, $x_I^+ x_I^-$ decreases, which in turn leads to the decrease of the dialton field too. Therefore, the dilaton field and the bulk CFT entropy decrease simultaneously when the boundary point $I$ moves outwards from the black hole horizon. In this case, we cannot obtain a reasonable solution for the boundary point $I$ outside the horizon. This can provide a physical reason for the discrepancy between our calculation and the one in Ref.\cite{Almheiri:2019yqk}.

\section{Conclusion and discussion}

For the eternal black holes, it is generally believed that the island extends outsides the horizon, i.e. the quantum extremal surface $I$ should be just outside the future event horizon. For the Liouville black holes, no reasonable island has been found in the expected region by extremizing the entanglement entropy.

\begin{figure}
  \centering
  \includegraphics[width=10cm]{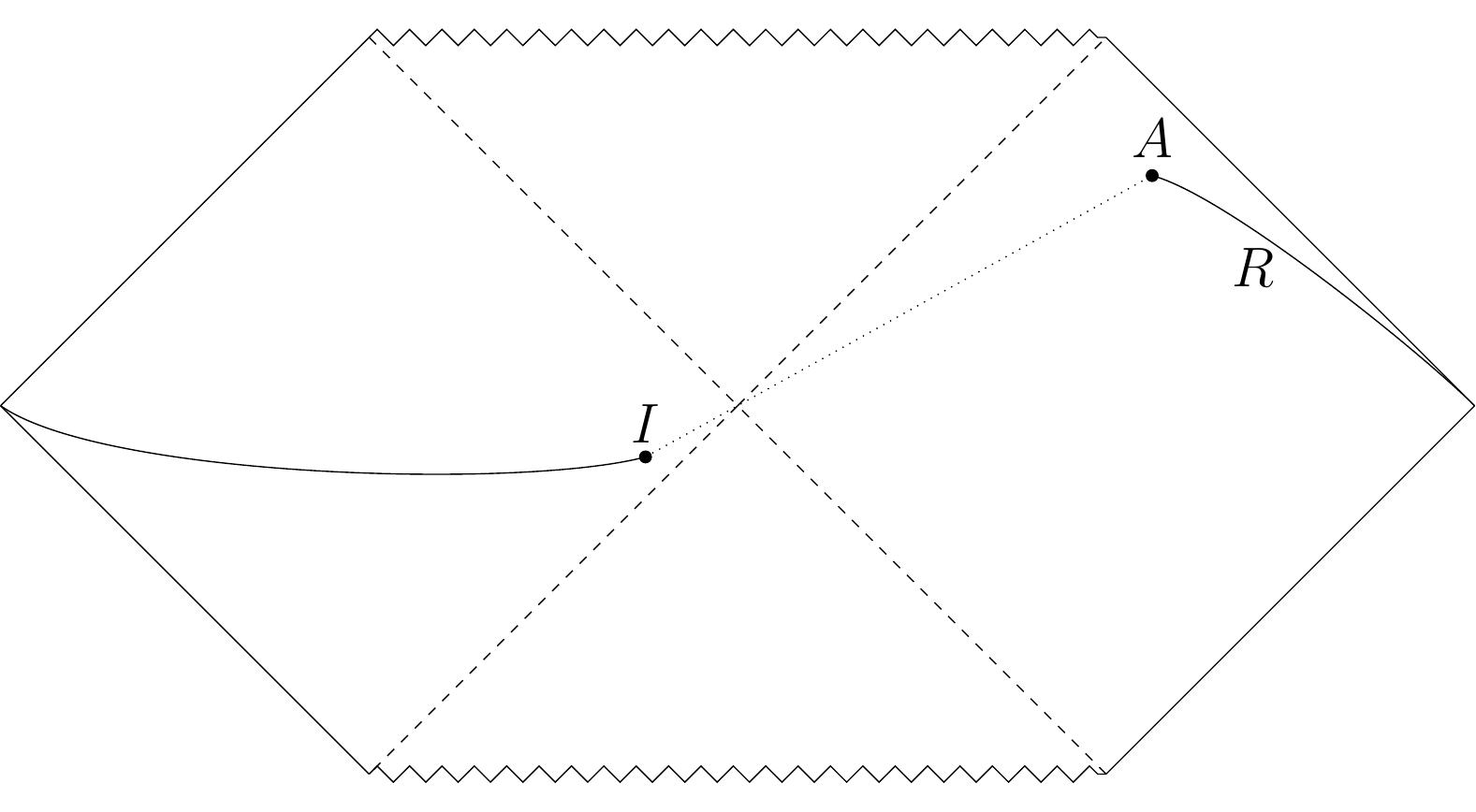}\\
  \caption{Penrose diagram with the novel island. It is assumed that the information paradox is experienced only by the observer in the right far from horizon region. }
  \label{PenroseDiagram_novelisland}
\end{figure}

One may imagine that the first group of solutions can be used to resolve the information paradox only experienced by the distant observer in the right region of the Penrose diagram. In this case, the entanglement wedge of the island is in the left region of the Penrose diagram as shown in Fig.(\ref{PenroseDiagram_novelisland}). This type of island was also found in pure de Sitter space \cite{Sybesma:2020fxg}. The location of the island in pure de Sitter space is also quite different from the cases when considering the black hole spacetimes. However, the island in de Sitter space can be properly interpreted as the requirement of no-cloning theorem. The novel island in the Liouville black holes seems to be unphysical.
However, if substituting this novel island solution in Eq.(\ref{novelsolution}) into the generalized entanglement entropy of radiation, we have the finite value at late time
\be\label{Snovel}
S\simeq S_{BH}+\frac{N}{12}\left(2\sqrt{|C|}b-\log(\lambda^2\beta C^2)\right)\;.
\ee
Notice that this entropy is half of the generalized entropy in Eq.(\ref{generalizedentropy}) because we only consider the entanglement entropy of the radiation in the right region.  
Although this novel island can reproduce the finite entanglement entropy of radiation, the scrambling time appears to be infinite. If we throw a diaries (strings of qubits) into the black hole from the right region, it will never reach the island. The Hayden-Preskill protocol \cite{Hayden:2007cs} can not be realized in this case. This means that the information swallowed by the black hole seems impossible to be released during the evaporation process.

In conclusion, we suggest that the island formula in the present form may not be used to resolve the information paradox for the two dimensional Liouville black holes. We also suspect the quantum extremal island formula depends strongly on the holographic duality and AdS spacetime. One should be very careful when applying the formula to study the information paradox of the asymptotically flat black holes.

\section*{Acknowledgement}

R.L. would like to thank Hongbao Zhang and Yuxuan Liu for helpful discussions.

 \end{document}